\documentclass{epl}
\usepackage{epsfig,subfigure}
\title{A non-Markovian optical signature for detecting entanglement in 
coupled excitonic qubits}
\author{F. J. Rodr\'{\i}guez\inst{1} \and L. Quiroga\inst{1} \and N. F. Johnson\inst{2}}

\institute{                    
  \inst{1}  Departamento de F\'{\i}sica, Universidad de Los
Andes, A.A. 4976, Bogot\'a, Colombia\\
  \inst{2} Physics Department, Clarendon Laboratory, Oxford University, Oxford OX1 3PU, U.K.
}
\begin{document}

\vspace{-4cm}
\maketitle

\begin{abstract}
We identify an optical signature for detecting entanglement in experimental nanostructure systems comprising
coupled excitonic qubits. This signature owes its strength to {\em non-Markovian} dynamical effects
in the second-order temporal coherence function of the emitted
radiation.  We calculate autocorrelation and
cross-correlation functions for both selective and collective light excitation, and prove that the
coherence properties of the emitted light do indeed carry information about the entanglement of the initial multi-qubit state.
 We also show that this signature can survive in the presence of a noisy environment.
\end{abstract}
\vspace{-0.95cm}
Essentially all the proposed schemes for quantum information
processing -- including quantum computing and teleportation --
require the reliable generation of entangled states between pairs
of qubits. Many physical implementations of qubits have been
proposed, and even built, in systems ranging from artificial
nanostructures through to naturally-occurring molecules and even
biological systems. However the following common problem faces {\em all} such
systems: Having gone through the effort of forming a set of
$N\geq 2$ qubits with the intention of undertaking some form of quantum
information processing, {\em how can we be sure that entangled
states are indeed being generated experimentally}?
More specifically, are there
any signatures for two-qubit systems that can distinguish between
states such as mixed or product states (which exhibit at most
classical correlations) and the crucially important entangled
states which carry purely quantum correlations? This question provides the first motivation for the present work. The second motivation stems from the fact that the
preparation and detection of highly-correlated quantum states are
difficult to perform in a controlled way due to the interaction of
the quantum system with the noisy environment -- after all, this is still one
of the main practical hurdles facing quantum computation. There is, therefore, a general need for
detailed quantitative theory concerning the time-evolution of
different initial quantum states, including their decoherence
properties.

In the present work, we propose such a signature, whose experimental
implementation can be achieved using current ultrafast optical
spectroscopy. In addition to this important practical finding, our
work provides a fundamental example of a quantum system where the
Markov approximation, which is invariably employed in such
calculations, yields incorrect results. Indeed, our results
suggest that any such optical probe of entangled states can {\em
only} be correctly interpreted using non-Markovian theories. The
Markov approximation, which is so convenient operationally, is
simply not good enough in these systems.
                                                                                                                                              
For the sake of concreteness, the qubit-matter systems we are
addressing can be realized using localized excitons in
semiconductor quantum dots (QDs)~\cite{lq,regelman,unold} or some
organic/biological molecules~\cite{hu} for which ultrafast optical
spectroscopic techniques are available. Moreover, charge or phase
superconducting qubits~\cite{nakamura} could also be studied within
the present proposal if the driving and detected radiation fields
are in the microwave sector of the electromagnetic spectrum.
Non-Markovian effects have been demonstrated in
CuCl $[6]$ and InGaAs$[7]$ QD's in a
picosecond time scale. At this short times the relaxation process
cannot be described by a single rate constant. For example, in
CuCl QDs the relaxation time is $\approx$ 2ps, thus the expected
value of the reservoir correlation time is or the order of at
least 1ps.
Additionally, ultrafast time-scales should be understood as femto-
or picoseconds for excitons, and microseconds for superconducting
qubits.

The photostatistics of photon counting as measured with the
correlation function $g^{(2)}(T,T+\tau)$ has triggered intense
theoretical and experimental activity in QDs
~\cite{rodriguez,sham,michler,gerardot} and biological systems
(chromophores)~\cite{zheng,mabuchi,weidemann}.
 This temporal second-order correlation function gives information about the conditional
probability of detecting one photon at time $T+\tau$ provided a
previous photon was found at time $T$.
 We are especially interested in the short time dynamics of such
correlations where the coupled system-environment could show some
unusual behaviour such as recoherence, among others. Therefore, a
proper description of the dynamics must be undertaken including
non-Markovian effects. Several theoretical approaches can be found
in the literature: the unraveling of non-Markovian stochastic
Schr\"odinger equation ~\cite{gambeta}, non-Markovian Monte Carlo
methods~\cite{dalibard} and the time-convolutionless projection
operator technique~\cite{breuer,emtage}. This last technique, based on the
Zwanzig projection approach~\cite{zwanzig}, has the advantage of
allowing the solution of a local-in-time master equation while taking
into account memory effects due to the system-environment
coupling. We adapt this last approach for the coupled qubit-noisy
environment which we address. We report three basic results: ({\sl i})
Unambiguous signatures of initial multi-qubit states;
({\sl ii}) detailed numerical calculations of the temporal
auto-correlation and cross-correlation, from which new
experimental features can be detected and analyzed at
ultrafast timescales, and ({\sl iii}) the identification of
induced effects by driving selectively an individual qubit, and its
effective non-classical radiation action on a second nearby qubit.

We consider the concrete example of two optically-driven,
dipole-dipole interacting qubits in contact with a boson bath. The
reduced density-operator in Lindblad form, is:
\begin{eqnarray}
i\frac{\partial \hat\rho}{\partial t}&=&\sum_{i=1}^{2}\Delta_i [S_i^z,\hat\rho]
+\sum_{i=1}^{2}\beta_i [S_i^\dag+S_i^-,\hat\rho]
+\sum_{i \neq j}^{2}V_{i,j}[S_i^\dag S_j^-,\hat\rho]
-i\sum_{i,j=1}^{2}\gamma_{i,j}(S_i^\dag S_j^- \hat\rho -
2 S_j^- \hat\rho S_i^\dag + \hat\rho S_i^\dag S_j^-)
\label{eq:1}
\end{eqnarray}
where $S_{i}^{\dag}$ is a qubit raising operator, $\beta_i$ is the
Rabi frequency and $\Delta_i=\omega_L-\omega_i$ the laser detuning
for the $i^{th}$-qubit. $\gamma_{i,j}$ and $V_{i,j}$ ($i \neq j$)
represent the collective decay and the dipole-dipole
interaction~\cite{varada} between qubits,
respectively. Using the Zwanzig\cite{zwanzig} projection technique, it is possible to
obtain a master equation governing the time evolution of the density operator
of the system in which the earliers time dependence in the density operator are 
included in a time dependent
decay rates.
The
non-Markovian effects are included by taking a time-dependent
spontaneous decay $\gamma_{i,i}(t)=\gamma_{i}(t)$\cite{breuer,zwanzig}.
In Eq.(\ref{eq:1}), where the first three terms
are associated with the coherent
evolution of the system and the last term describes decoherence
processes,
there are three control parameters: the laser field
strength $\beta_i$, the driving laser detuning with the qubits,
and the inter-qubit separation $r_{12}$ which affects
$V_{i,j}$.  We assume that the propagation direction of the incident
field is perpendicular to the interqubit axis, $\vec k \perp \vec {r}_{12}$.
Note that although our theory employs a generalization of Markovian decay-rates
to non-Markovian situations, i.e. $\gamma_{ij}(t)=\gamma_i(t)$, the strong
resulting enhancement of photon-photon correlation effects which we observe
is a {\em highly non-trivial} consequence of this generalized decay rate.
In addition, a wide range of monotonic time-dependent functions $\gamma_i(t)$
should show similar effects (but different from the $\gamma$ constant case),
making our results insensitive to the precise form assumed for $\gamma_{ij}(t)$\cite{diosi,saikan}.

An important aspect of the present work is the analysis, at very short times, of the finite memory response of
the reservoir ($\gamma_r^{-1}$) due to the qubit-boson environment
interaction\cite{diosi}. Extending the projection technique result
~\cite{rodriguez} to the present coupled qubit case, the qubit
spontaneous decay is taken as $\gamma_i(t)=\Gamma_0 (1-e^{-\gamma_r
t})$, where $\Gamma_0$ is the usual (stationary or Markov)
decoherence decay constant.
In order to preserve the positivity of the
reduced density matrix, we consider an intermediate regime of
coupling strengths between the environment and the qubits.
                                                                                                                                              
We focus on three central
optical measurements: {\sl (i)} the autocorrelation function
($i=j$), {\sl (ii)} the cross-correlation functions ($i\neq j$)
given by 
\begin{eqnarray}
g^{(2)}_{i,j}(T,T+\tau)=
\frac{
(\langle S_j^\dag (T) S_i^\dag (T+\tau) S_i^- (T+\tau) S_j^-(T)\rangle)
}
{
(
\langle
S_j^\dag (T)S_j^-(T)\rangle \langle S_i^\dag (T+\tau) S_i^-
(T+\tau)\rangle
)
}
\end{eqnarray}
Physically, these functions represent the
conditional probability of detecting one photon emitted from the
$j^{th}$ qubit at time $T+\tau$ after detecting one photon emitted
from the $i^{th}$ qubit at $T$; their importance relies on the
fact that they are independent of the propagation observation
angle and also that they are easier to obtain experimentally.
Finally, {\sl (iii)} the full photon correlation is given by
\begin{eqnarray}
g^{(2)}(T,T+\tau)=\frac{\sum_{i,j,l,m}\langle S_i^\dag (T)
S_j^\dag (T+\tau) S_l^- (T+\tau) S_m^-(T)\rangle e^{i {\vec
k}.{\vec r}_{i,j}}e^{-i {\vec k}.{\vec r}_{l,m}}} {\sum_{i,j}
\langle S_i^\dag (T)S_j^-(T)\rangle e^{i {\vec k}.{\vec r}_{i,j}}
\sum_{l,m} \langle S_l^\dag (T+\tau) S_m^- (T+\tau)\rangle e^{i
{\vec k}.{\vec r}_{l,m}}} 
\end{eqnarray}
where ${\vec k}$ has magnitude $k=
(\omega_1+\omega_2)/2c$ and its direction coincides with the
far-field observation direction ($i,j,l,m=1,2$).

As a first step for calculating $g^{(2)}$, we numerically solve
the reduced density matrix equation to describe the time-evolution
of qubit correlations such as $\langle
S_i^\dag(T)S_j^\dag(T+\tau)S_k^-(T+\tau)S_l^-(T)\rangle$. Since we
describe the interacting qubit-environment dynamics as a
non-Markovian process, it turns out that the set of equations to be
solved closes in a finite hierarchy of operator correlations. The above correlation couples,
in a nontrivial manner, with higher-order correlations of type
$\langle S_i^{\pm}(T)S_j^{\pm}(T+\tau)S_k^{\pm}(T+\tau)S_l^{\pm}
(T+\tau)S_m^{\pm}(T+\tau)S_n^{\pm}(T)\rangle$,
due to the fact that the
quantum regression theorem cannot be safely used in
the present context where memory effects are included.
                                                                                                                                              
We
consider two important experimental setups:{\sl (i)  a collective
excitation} where both
qubits are simultaneously illuminated with the same non-resonant laser
intensity, and {\sl (ii) a selective excitation} scheme where only
one qubit is driven by the laser producing non-classical light,
which in turn can excite the second qubit or can be registered by
the detector. In the first case, it is impossible to distinguish
the light emitted from each qubit while in the second case
anti-correlation effects can be observed indicating the origin of
the detected emitted photon.
Additionally, different initial qubit states are analyzed:
(i) A separate (non-entangled) state formed by the
product of identical superposition states for each qubit, $\mid
\Psi_{QQ} \rangle=\frac{1}{2}(\mid 0_1 \rangle + \mid 1_1
\rangle)\otimes(\mid 0_2 \rangle + \mid 1_2 \rangle)$, where the
\begin{figure}[htbp]
\centerline{
{\includegraphics [height=5.0cm,width=5.8cm]{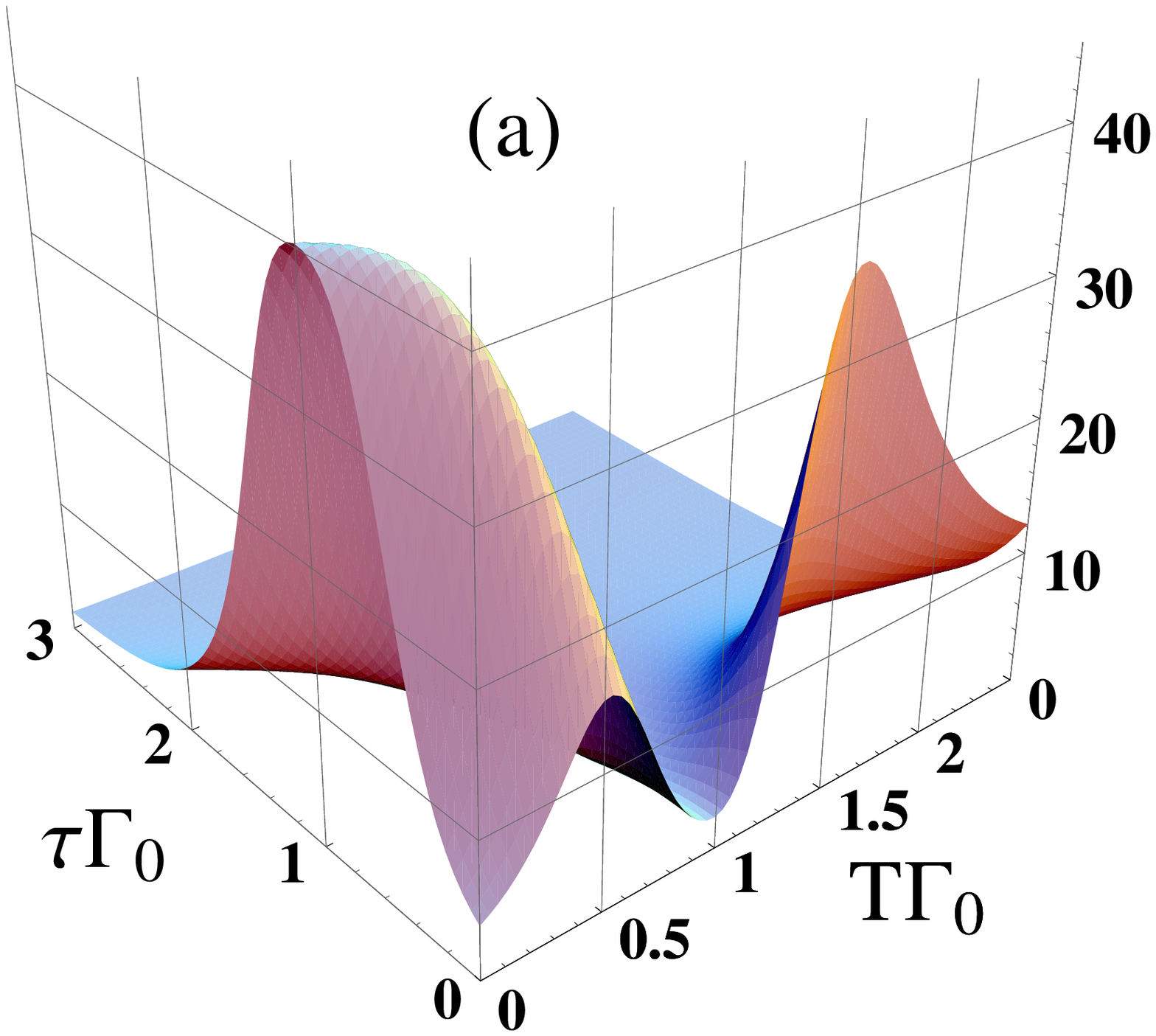}}
\hspace{2cm}
{\includegraphics [height=5.0cm,width=5.8cm]{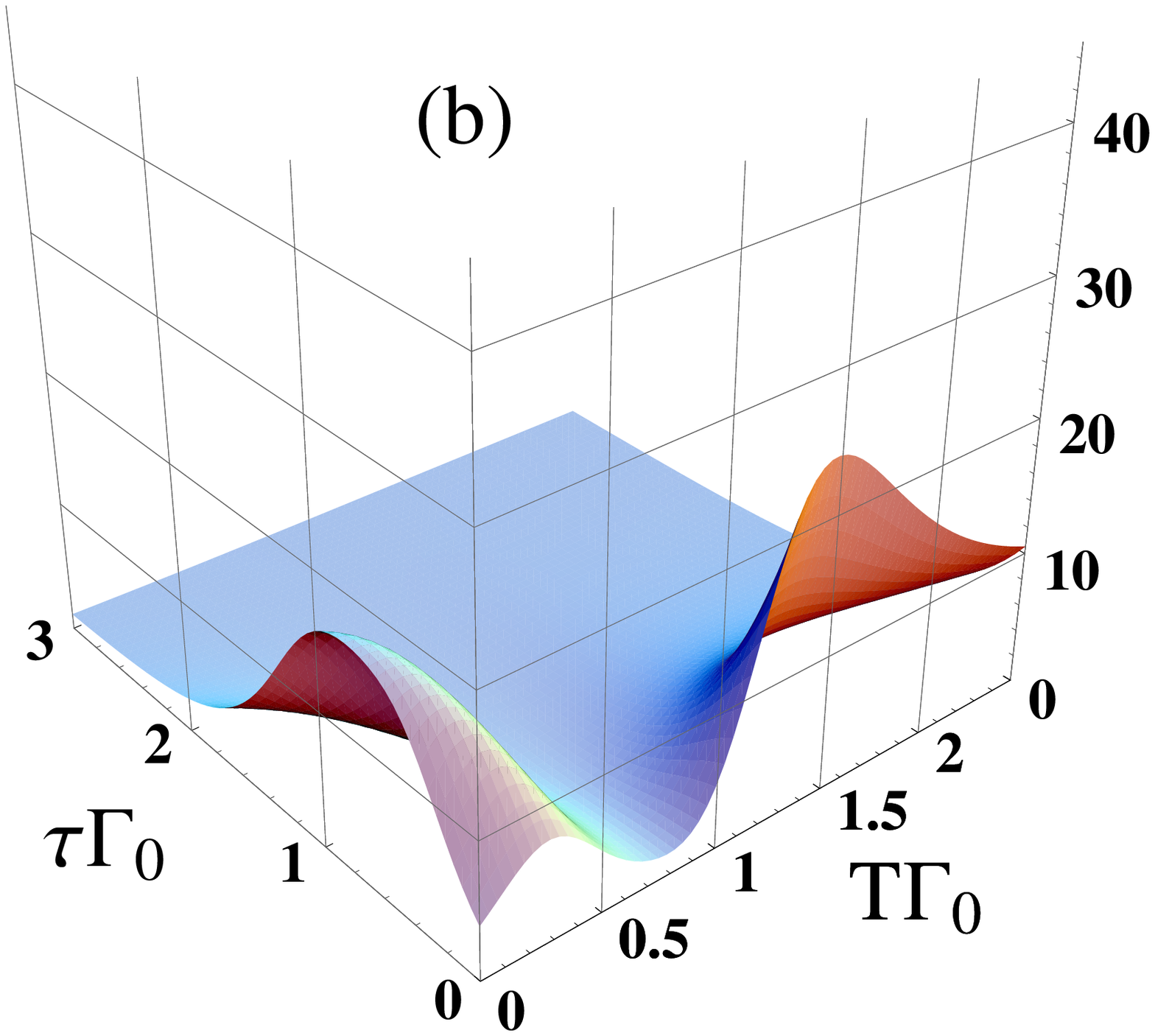}}
} 
\caption{
The full correlation function
$g^{2}(T,\tau)$ for a separable initial state ($|\Psi_{QQ}\rangle$) with
$\beta= 1.25\Gamma_0$. (a) Non-Markovian case with $\gamma_r=2\Gamma_0$, it
means that the reservoir correlation time $\tau_r=\hbar/2\Gamma_0=16 ps$ and
(b) Markovian case ($k r_{12}=\pi$).}
\label{fig-1}
\end{figure}
\vspace{-0.01cm}
subscripts indicate the $i=1,2$ qubit. (ii) A maximally entangled
Bell state $\mid \Psi_{B} \rangle=\frac{1}{\sqrt{2}}(\mid 0_10_2
\rangle + \mid 1_11_2 \rangle)$. (iii) An entangled triplet state
$\mid \Psi_{T} \rangle=\frac{1}{\sqrt{2}}(\mid 0_11_2 \rangle +
\mid 1_10_2 \rangle)$. (iv) An entangled singlet state
(EPR-state) $\mid \Psi_{S} \rangle=\frac{1}{\sqrt{2}}(\mid 0_11_2
\rangle - \mid 1_10_2 \rangle)$.

{\it Collective excitation ($\beta_1=\beta_2=\beta$).-} In Fig.1, the full correlation
function $g^{2}(T,\tau)$ is plotted for an initial separate state
$|\Psi_{QQ} \rangle$.
Both qubits are simultaneously
illuminated with a laser excitation of 1.25$\Gamma_0$
($\Gamma_0$=20 $\mu eV$).
It is possible to estimate the dipole moment by $\mu E\pi
\tau_p/\hbar=A$, where $E_0$ is the radiation electric field,
$\tau_p\approx 6 ps$ is the window time of the incident pulse and
$A$ is the pulse area of 0.24$\pi$\cite{andreani}, which corresponds to a typical
experimental excitation (with $\mu\approx 8$
Debyes)~\cite{regelman}. The central laser frequency is detuned
$\Delta_1=-0.5\Gamma_0=-\Delta$ and $\Delta_2=0.5\Gamma_0=\Delta$, from each qubit.
A strong antibunching of the
emitted radiation is evident at very short times. Note that, for
$0<T<\Gamma_0^{-1}$, the second-order correlation function is
rather immune to decoherence effects in the non-Markovian case
(Fig.\ \ref{fig-1}.(a)) as compared with the Markovian case
(Fig.\ \ref{fig-1}.(b)). This is a consequence of the
{\em recoherence} processes between the environment and the two
 qubits. The system decoheres,
interacting with the bath, and at $\tau\Gamma_0\approx 1$ a
correlation revival arises. This strong antibunching effect is a
non-Markovian process. Thus, the quantum correlations at time $T+\tau$
(a first photon detected at $T$) are affected by the type of initial state
 at time $T=0$. This effect
is enhanced for individual qubit superposition states, and is a
consequence of the vanishing values taken by the intensity
correlations $\sum_{l,m} \langle S_l^\dag (T+\tau) S_m^- (T+\tau)\rangle$.
We emphasize that this effect is stronger for two interacting qubits as a function
of $\tau$,
 as compared with the single qubit case previously reported in Ref.~\cite{rodriguez} and persists
for longer $T$ times.
This can be understood by analyzing the initial state as: $\mid
\Psi_{QQ}
\rangle=\frac{1}{2}(|1_1,1_2\rangle+|0_1,0_2\rangle+|0_1,1_2\rangle+|1_1,0_2\rangle)$.
This state comprises the superposition of one exciton in each
qubit, one non-radiative state (zero excitons) and the triplet
state (which is associated to a superradiant state). The
contribution from the triplet state provides a distinctive
signature which is significantly enhanced by non-Markovian effects. For long times
($T\Gamma_0=2$) an additional peak appears in the second-order
correlation function, as a consequence of the typical
oscillations given by the external parameters at
$T\Gamma_0=4(\Delta^2+(2\beta)^2+(V_{12}-\Gamma_0)^2)^{-1/2}$.
                                                                                                                                  
\begin{figure}[htbp]
\centerline{
{\includegraphics [height=5.0cm,width=5.8cm]{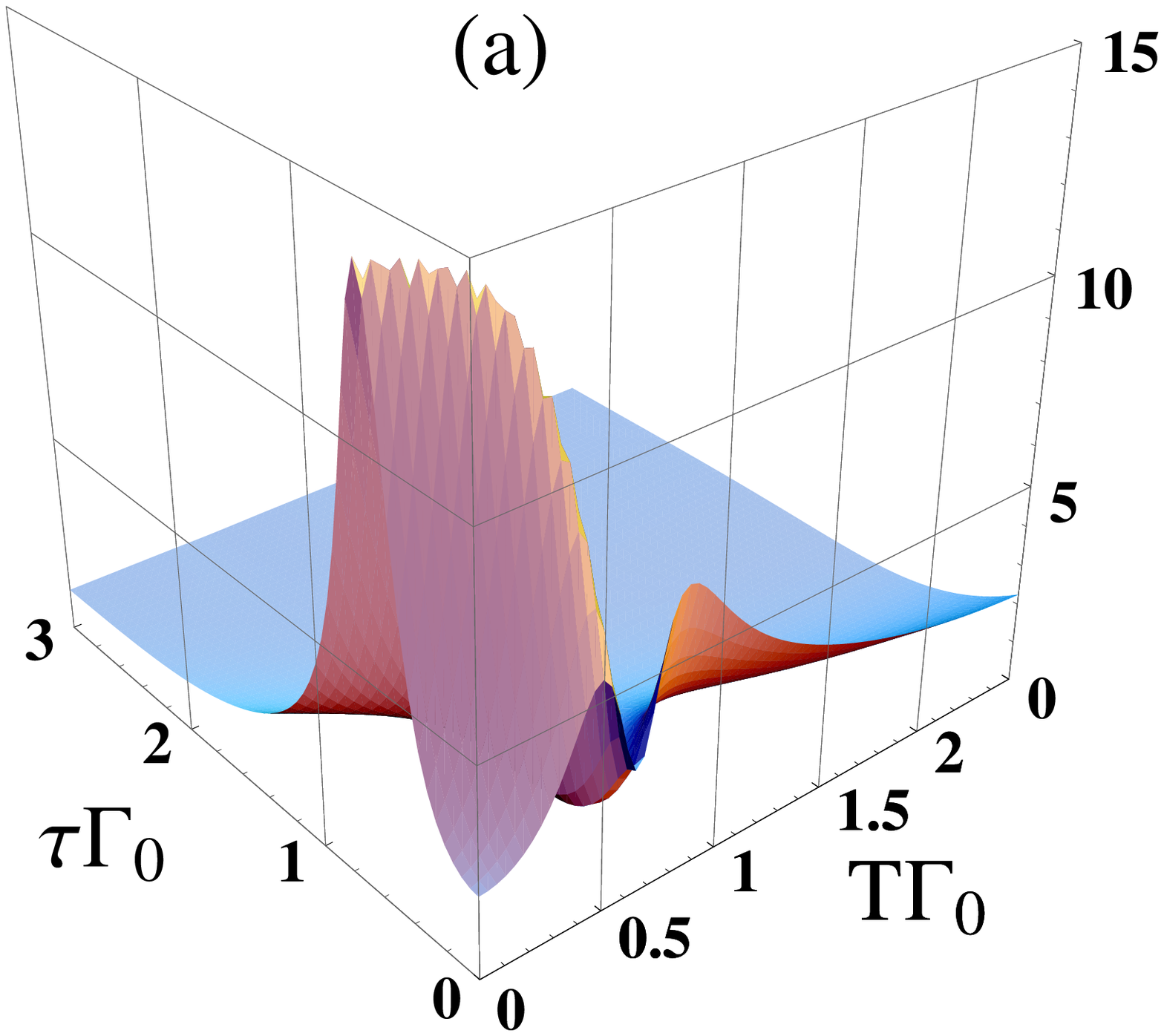}}
\hspace{2cm}
{\includegraphics [height=5.0cm,width=5.8cm]{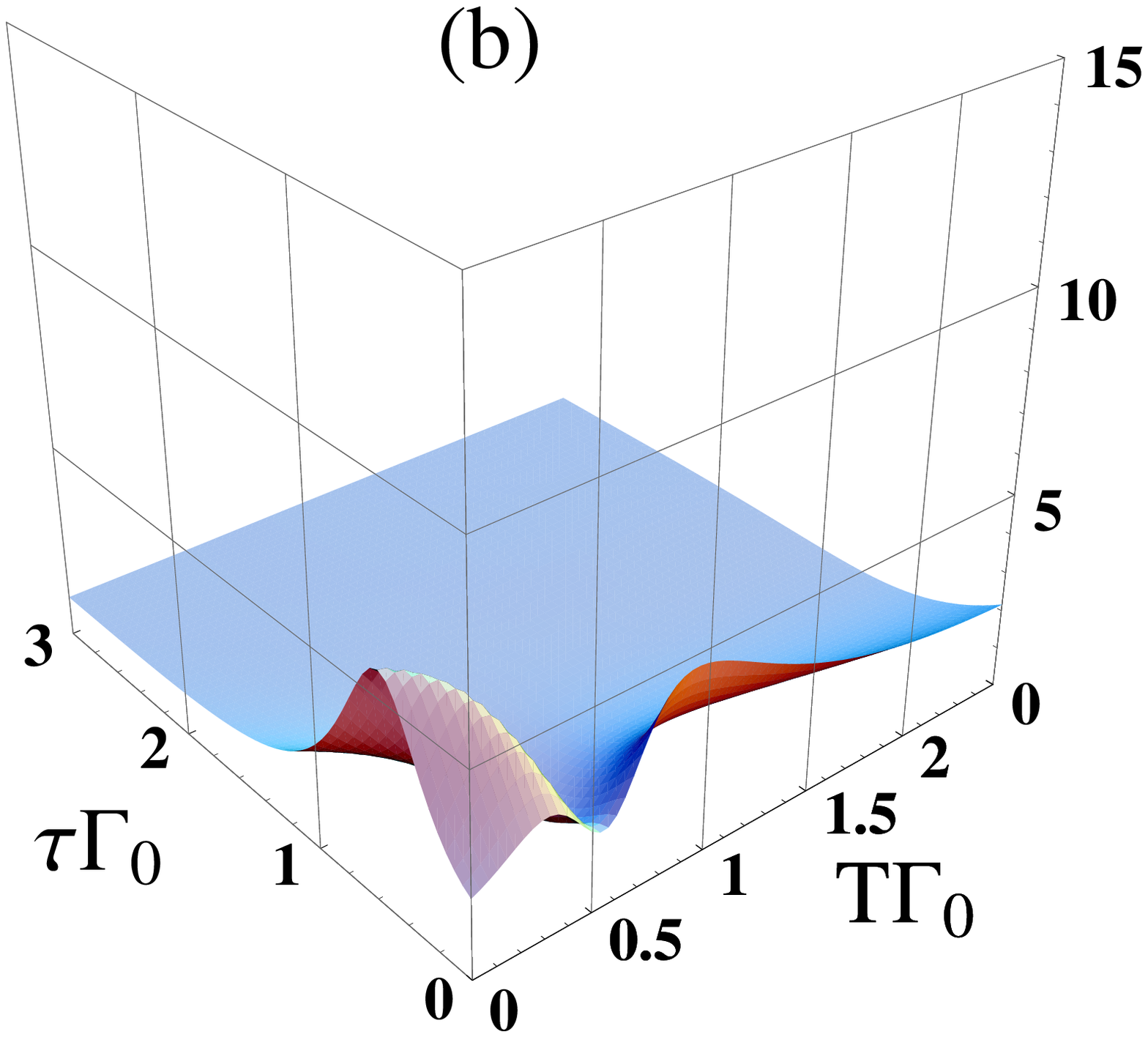}} }
\caption{
 The full correlation function
$g^{2}(T,\tau)$ for an initial Bell state. (a) Non-Markovian case
and (b) Markovian case. Parameters as in Fig.\ (\ref{fig-1}) but
now with a common Rabi frequency $\beta=2.25\Gamma_0$.}
\label{fig-2}
\end{figure}
\begin{figure}[htbp]
\centerline{
{\includegraphics [height=5.0cm,width=5.8cm]{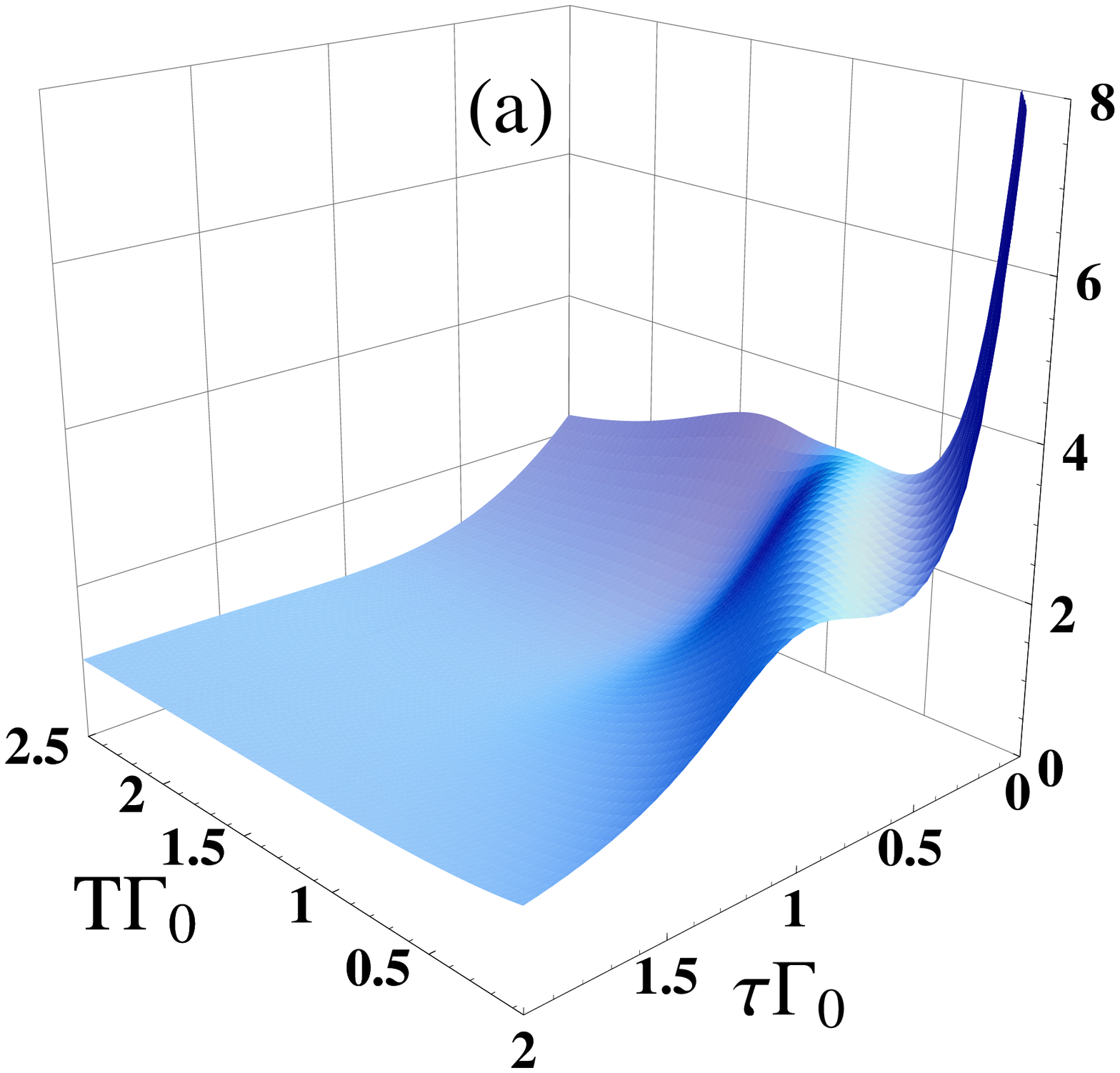}}
\hspace{2cm}
{\includegraphics [height=5.0cm,width=5.8cm]{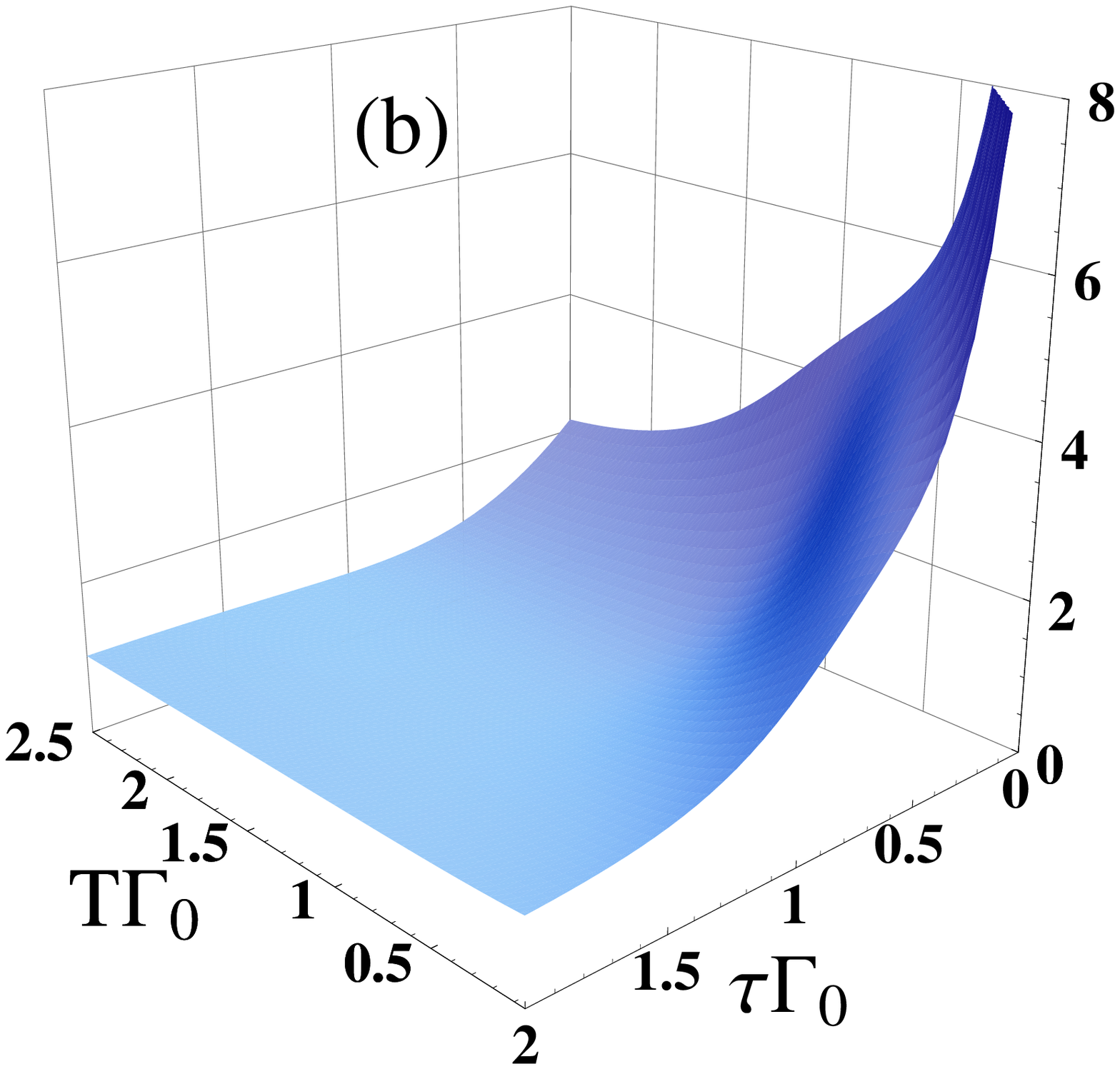}}
}
\caption{
The complete correlation function $g^{2}(T,\tau)$ in the vertical axis for the excitonic triplet state.
(a) Non-Markovian case and (b) Markovian case.
Antibunching is comparable with a superposition state.}
\label{fig-3}
\end{figure}

\vspace{-0.01cm}
In order to stress the quite different optical signatures for
different possible initial states, Fig.\ \ref{fig-2}(a) focuses on an
initial Bell state: $\mid \Psi_{B} \rangle$. This state does not
contain the contribution of the triplet state which explains why
the intensity of the corresponding correlation function is about
three times smaller than the one corresponding to the
separate state $\mid \Psi_{QQ} \rangle$. However, the
non-Markovian effects can still be observed, when compared with
the results of Fig.\ \ref{fig-2}(b) for the Markovian case. The
origin of this strong antibunching signature can be understood by
inspecting Fig.\ \ref{fig-3}, with the same parameters as Fig.\
\ref{fig-1} and \ \ref{fig-2}, which shows the results for the
triplet state: $\mid \Psi_{T} \rangle$. The $g^{2}(T,\tau)$
results yield a huge (one order of magnitude higher)
antibunching effect at very short $T$ times in clear contrast to
the Bell state. The oscillations remain in the non-Markovian case, as
compared with the Markovian case (Fig.\ \ref{fig-3}(b)). Our results clearly
show one distinctive signature: the triplet state is
responsible for a strong antibunching effect, which is
enhanced due to the recoherence processes between the environment
and the qubit system. This effect can be tailored and also
enhanced if the inter-qubit separation is reduced. We emphasize that our
parameter values are consistent with current
experimental data, and the results presented can be produced with low laser
excitation provided that $\beta\geq \gamma_r$.

Summarizing, at very short times the slope
of $g^{2}(T,\tau)$ is positive indicating an antibunching
behaviour, which is enhanced by non-Markovian effects
in the range of $0<\tau\Gamma_0<1$ and
$0<T\Gamma_0<1$, allowing the distinguishability among the $\mid \Psi_{QQ} \rangle$,$\mid \Psi_{B}
\rangle$ and $\mid \Psi_{T} \rangle$ states. As can be seen in Fig \ \ref{fig-3} the triplet
state lacks robustness to decoherence effects. The EPR-state ($\mid \Psi_{S} \rangle$),
does not show significative differences in its time-evolution when memory
effects are taking into account (not shown).
 It is also interesting that the
second-order time correlations reach higher values for separable states
but {\em decrease} when the input state becomes entangled.
 Finally, for $T\Gamma_0>2$ the
non-Markovian results tend toward the Markovian ones, emphasizing that in this long time regime
the time-correlation is stationary and
independent of the initial states.

\begin{figure}[htbp]
\centerline{
{\includegraphics [height=4.9cm,width=9.0cm]{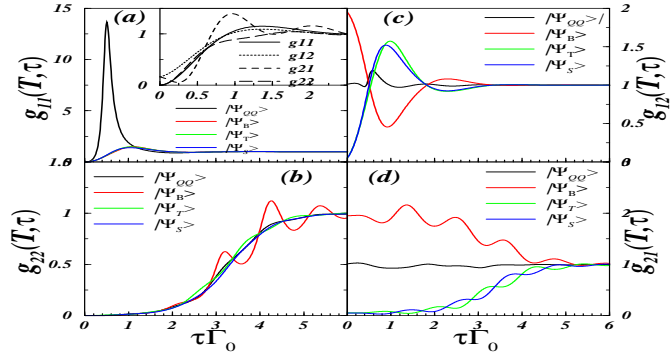}}
}
\caption{
(on-line color). Selective escitation with a Rabi frequency of $\beta_1=2.25\Gamma_0$. Interqubit separation as in Fig. \ (\ref{fig-1}).
 (a) and (b) show the auto-correlated functions and
(c) and (d) are the cross-correlated ones.
The different signals correspond to a superposition (black-line), Bell (red-line),
symmetric (green-line), anti-symmetric (blue) states. The inset depicts the stationary limit.}
\label{fig-4}
\end{figure}

{\it Selective excitation ($\beta_1=\beta,\beta_2=0$).-} We now consider the case when only
one qubit is externally driven by a laser light and each qubit is
out of resonance. In this scheme, it is possible to obtain
information about the signatures of the light emitted from a
second qubit when antibunched light emitted from the excited qubit
is driving the second qubit. In Fig.\ \ref{fig-4}, we show
comparative results for different initial states ($T\Gamma_0=1$),
where the non-Markovian signatures can still be observed. In Fig.\
\ref{fig-4}(a) and (b), the autocorrelation functions are
completely different, due to the fact that the non-illuminated
qubit does not emit photons. Once the non-driven qubit becomes
excited, $g_{11}^{(2)}(\tau)$ decreases while
$g_{22}^{(2)}(\tau)$ increases. However it is interesting to note
that both autocorrelations vanish at $\tau=0$, but start to
increase towards unity for long times.
This leads us to consider the cross-correlations
$g_{12}^{(2)}(\tau)$ and $g_{21}^{(2)}(\tau)$, which allow to know if the
information about the emitted correlated photons and if the
emitted photons are uncorrelated ($g_{i,j}^{(2)}(\tau)\leq 1$) or
correlated (for $i\neq j$).
 The differences are clearly visible in Fig. 4(c).
For separate state ($\mid \Psi_{QQ} \rangle$) the plot shows not
difference between $g_{12}$ and $g_{21}$ which correspond to a complete
  uncorrelated situation for any $\tau$.
The $\mid \Psi_{B} \rangle$
state, exhibits strong correlations for $\tau\Gamma_0 \leq 0.5$
($g_{12}\geq 2$) with a time dependent correlations going to zero.
However, by selectively
exciting just one qubit, this strong correlation remains for
longer times (up to $\tau\Gamma_0 \approx 2$, see Fig.4(d)).
The anticorrelation is perfect for the
entangled triplet and singlet states, indicating that there are no
photons available to excite the second qubit and thus it is
possible to distinguish the source of the detected photons. This
effect can also be observed in the stationary regime ($T\rightarrow\infty$), as is
clearly seen in the inset of Fig.4(a) for times $\tau\Gamma_0 < 0.5.$
However, the memory is
lost due to the interaction with the environment and therefore
no information about the initial state is preserved. The
cross-correlation, $g_{21}$ goes to zero at a very
short time ($\tau\Gamma_0 \approx 0.2.$) in contrast with
the non-Markovain case where for a long period of time, the
cross-correlations remain almost at zero value for the entangled
singlet and triplet qubit states.

To summarize, we have shown that for an optically-driven
coupled qubit system, differently prepared initial states -- for example, separate
superpositions, Bell and triplet (symmetric) states -- can be clearly
distinguished as a direct result of non-Markovian effects. The calculation of
normalized autocorrelation and cross-correlation second-order
functions, also give valuable information concerning the extent to which each qubit
behaves like a single quantum light source. We have demonstrated
how it is possible to assess the entanglement content
by observing the correlations function of the emitted light
for collective as well as selective excitations. In the latter case,
the cross correlations give enough information about the photon source
by exploiting the non-classical light produced by a qubit on a nearby
second single qubit.
 This highly correlated light could be used to probe systems sensitive to non-Markovian effects.
Our study exposes new
phenomena that can be observable at ultrafast scales of time for which
non-Markovian dynamics should be dominant.
To our knowledge, the proposed signature is the first of its kind and should act as a
powerful tool for experimental groups attempting to realize quantum information processing
in a wide range of optically-driven, coupled-qubit systems.
\vspace{-0.61cm}
\acknowledgments
The authors acknowledge partial support from COLCIENCIAS (Colombia) projects No. 1204-05-11408, 1204-05-13614, 
Facultad de Ciencias (2005-2006) and Banco de la
Rep\'ublica (Colombia).

\end{document}